\documentstyle[aps,epsfig,prb]{revtex}

\def\ee {\text{e}}
\def\dd {\text{d}}
\def\e {\epsilon}
\def\eps {\varepsilon}
\def\o {\omega}
\def\D {{\cal D}}
\def\d {{\Delta}}
\def\Sp{\textsf{Sp}\,}
\newcommand\U{\textsf{U}\,}
\newcommand\SO{\textsf{SO}\,}
\def\S{\textsf{S}\,}
\def\Tr {\text{Tr}\,}
\def\tr {\text{tr}\,}

\def\l {{\lambda}}
\def\L {{\Lambda}}

\begin{document}
\def\theequation{\arabic{section}.\arabic{equation}}%

\title{Non-perturbative results for level correlations
from the replica  nonlinear $\sigma$ model}
\author{ I. V. Yurkevich and I. V. Lerner}
\address{School of Physics and Astronomy, The University of Birmingham,
Edgbaston, Birmingham B15 2TT, UK}

\date{\today}
\maketitle

\begin{abstract}
We show that for all the three standard symmetry
classes  (unitary, orthogonal and symplectic), 
the conventional 
 replica  nonlinear $\sigma$ model gives 
the correct non-perturbative result for the 
two-level correlation functions $R_2(\o )$ of electrons in disordered metals 
in the limit of large $\o $.
In this limit, non-perturbative oscillatory contributions
arise from a degenerate saddle-point 
manifold within this  $\sigma$ model which corresponds to the 
replica-symmetry breaking. 
Moreover, we demonstrate that in the unitary case the very same results
can be extracted from the well known exact integral 
representation for $R_2(\o )$.
\end{abstract}

\bibliographystyle{simpl3}

\section{Introduction}\setcounter{equation}{0}

Starting from the seminal papers of Wegner\cite{Weg:79} and Efetov,\cite{Ef:82} 
a field-theoretical description based on the nonlinear $\sigma$ 
model (NL$\sigma$M) has become one of the main analytical approaches to 
various problems in disordered electronic systems. 
The ensemble averaging over all configurations of disorder is performed
either using bosonic\cite{Weg:79,Sch+W}
or fermionic\cite{EfLKh} $n$-replicated fields and taking the 
$n\!\to\!0$ limit in the results,  
or using supersymmetric ($Z_2$-graded) fields.\cite{Ef:82,VWZ}

The very first application of this approach was a derivation 
\cite{Sch+W,EfLKh,Ef:82} of the renormalization-group (RG) equations
of the scaling theory\cite{AALR} of Anderson
localization. For such a perturbative derivation, 
generalized later for
mesoscopic systems\cite{AKL:86}, both the replica and the supersymmetric
methods are equally well justified: they just ensure the cancellation 
of unphysical vacuum loops in a diagrammatic expansion.
However, it soon became conventional wisdom
that there existed two sets of problems for which  
only one of these methods was applicable.

On the one hand, the fermionic replica NL$\sigma$M 
has been generalized by Finkelstein\cite{Fin}
to include interactions between electrons. The interest in this 
approach has been greatly enhanced by the recent discovery\cite{Krav:95}
of an apparent metal-insulator transition in 2D disordered systems
in zero magnetic field. Although it has been recently 
demonstrated\cite{AnKam:98}
 that the Keldysh technique provides a viable alternative to the 
replica approach, the latter still remains one of the best available
tools for consideration of interacting electrons in disordered systems. 
In the very least, it is clear that there is no simple way
of applying the supersymmetry method to a
many-particle {\it fermionic} system. 

On the other hand, the viability of the replica approach was undermined
by the existence of a set of problems which could apparently be
solved only with the help of the supersymmetry method. 
The first, and possibly most famous, of such problems was
solved by Efetov\cite{Ef:82b} who used the supersymmetric
NL$\sigma$M to derive the two-level correlation function (TLCF)
in the universal ergodic regime for electrons in disordered metallic grains. 
The results proved to be identical to those for eigenvalue correlations
in random matrix theory,\cite{RMT} as had much earlier been assumed by
Gor'kov and Eliashberg.\cite{GE}
This first microscopic derivation of essentially non-perturbative results
has opened the way to numerous new results (mainly obtained
during the last decade) 
 for which using the supersymmetry
method seemed to be absolutely essential (many of these results have
been reviewed in a recent book by Efetov\cite{Efetbook}). 

In many ways, though, this first non-trivial ``supersymmetric''
result\cite{Ef:82b}
 seemed to be the best illustration of why the replica method 
could only be used within a perturbative approach. 
For the easiest case of the unitary symmetry
 the irreducible TLCF is given by
\begin{equation}
\label{R2U}
R_2(\o ) = -\frac{\sin^2\!\o }{\o ^2}\,,
\end{equation}
where $\o $ is the distance between two levels in units of
$\d /\pi$ and $\d $ is the mean level spacing. This result is valid
in the ergodic
regime, i.e.\ for $\o \ll g$ where $g\gg1$ is the dimensionless conductance.
For $\o \gg 1$, the TLCF
averaged over fast oscillations, could be readily obtained from 
the standard diagrammatic techniques \cite{AS} with $1/\o $
being the perturbation parameter (or from the perturbation
theory in the framework of either the supersymmetric or replica NL$\sigma$M).
However, the non-perturbative factor
$\sin^2\! \o $ cannot be restored from the perturbation
series. Since the replica trick is well justified only within 
the perturbative approach, it might seem rather hopeless
to obtain the result (\ref{R2U}) within the replica approach. 
And indeed, quite involved calculations by 
Verbaarschot and Zirnbauer \cite{VZ:85} have shown that 
a direct application of the replica trick (using either the bosonic or
fermionic NL$\sigma$M) has apparently  not reproduced
 the TLCF given by Eq.~(\ref{R2U}).

Therefore, a very recent result of Kamenev and Mezard \cite{KamMez:99a}
makes a real breakthrough in this area. Using the replica-symmetry
breaking 
within an effective fermionic field theory corresponding to
the large $N$ limit
of the Gaussian unitary ensemble of $N\times N$
random matrices,  they have reproduced the non-perturbative
result (\ref{R2U}), albeit only in the region $\o \gg1$. 
This raises hope that one might eventually apply the replica approach
for obtaining {\it non-perturbative results} for 
interacting electrons in disordered systems. 
However, the method used in Ref.\ \onlinecite{KamMez:99a}
does not appear to be extendible to the standard 
NL$\sigma$M describing  electrons in a random potential.
 Although the large-$N$ field
 theory written in terms of the $n\times n$ replica 
matrix fields $Q$ is, in principle, 
totally analogous to the  standard 
NL$\sigma$M, this analogy has been considerably smeared by the very essence
of the method applied. Namely,
the standard $\exp [N\tr\ln(\e -i Q)]$ term has been represented after
a shift of variables  as $(\det Q)^N$  and the resulting Itzykson-Zuber
integral (similar to that 
introduced in the context of the supersymmetric method 
by Guhr\cite{Guhr})
was calculated taking into account replica-symmetry breaking.
\cite{KamMez:99a}

Although the method used in Ref.\ \onlinecite{KamMez:99a}
could hardly be generalized directly, the message was very clear: the replica
approach taken together with the symmetry-breaking, can reproduce the
results  of the supersymmetric approach. 

The purpose of this paper is to demonstrate that non-perturbative
results for all the classical symmetry ensembles
(unitary, orthogonal and symplectic)
can be derived in the framework of the standard fermionic
NL$\sigma$M with replica-symmetry breaking. We believe that the present
derivation is both more general and considerably easier than the original one
\cite{KamMez:99a}
and can be straightforwardly extended to include interactions. 
Independently, Alex Kamenev and Marc Mezard have presented
in their second paper \cite{KamMez:99b} a derivation which is 
similar in spirit, although considerably different from that suggested here.

Before describing our derivation, we want to emphasize the following. 
It is the standard fermionic NL$\sigma$M which does contain correct
(at least for large $\o $) non-perturbative 
oscillatory contributions to the level correlation
functions. The replica-symmetry breaking which involves a set of additional
saddle-point submanifolds within the standard NL$\sigma$M 
is just a very convenient method of calculating the large-$\o $
asymptotic behavior of the correlation functions including these
oscillatory contributions. Similarly, within the supersymmetric NL$\sigma$M
Andreev and Altshuler \cite{A2} have reproduced 
the large-$\o $ limit of Eq.\ (\ref{R2U}) by the saddle-point calculation
of the supersymmetric integral that involved an additional 
 saddle-point which breaks the supersymmetry. The very same
integral has been exactly calculated by Efetov\cite{Ef:82b} 
without breaking the supersymmetry. We extend this analogy to the fermionic
replica NL$\sigma$M by showing, with the help 
of a trick which resembles the replica-symmetry breaking,
 that the $n=0$ limit of the exact 
integral representation  of the TLCF obtained
 by Verbaarschot and Zirnbauer \cite{VZ:85} does contain, at least for
large $\o $,
the correct oscillatory behavior.

The paper is organised as follows. 
In the next section we recall the formulation of the standard
replica fermionic  NL$\sigma$M and introduce all relevant notations.
In Section \ref{sec:unitary}
 we present in  detail a calculation of the  large $\o $ non-perturbative
contribution to the TLCF for the unitary ensemble, based on 
replica-symmetry breaking. In Section  \ref{sec:ort}
 we outline similar calculations
for the orthogonal and symplectic ensembles. In Section \ref{sec:VZ}
 we extract  the very same results 
 from the exact integral representation for $R_2$ obtained by Verbaaschot and
Zirnbauer \cite{VZ:85} for the unitary ensemble. Finally, in 
Section \ref{sec:con} we summarize
our results and discuss perspectives of the replica method.

\section{Formulation of the problem}\setcounter{equation}{0}

We consider the two-level correlation function (TLCF) defined by
\begin{equation}
\label{TLCF}
R_2(\o ) = \frac1{\nu^2}\left<\nu(\eps + \o ) \nu( \eps )\right>-1\,.
\end{equation}
Here $\left<\ldots\right>$ stands for the ensemble averaging, $\nu(\eps )$
is the electronic density of states per unit volume defined in terms 
of the spectrum $\{\eps _\alpha\}$ for a given sample as $\nu(\eps )
= L^{-d} \sum_\alpha\delta(\eps - \eps _\alpha)$, $\quad\nu\equiv
\left<\nu(\eps ) \right> =1/L^d \d $, and $L$, $d$ are the sample size and
dimensionality. We will measure all energies in units of $\Delta/\pi$,
in which the TLCF is as written in Eq.~(\ref{R2U}). In these
units we express $R_2$ via the product of the retarded and advanced
Green's functions as follows:
\begin{equation}
\label{S2}
R_2(\o)=\frac{1}{2}\left[\Re\text{e}\, S_2(\o)-1\right]\,,
\qquad  S_2(\o)\equiv \left<  G^r\left(\eps +\frac{\o}{2}\right)
G^a(\eps -\frac{\o}{2})
\right>\,.
\end{equation}
The function $S_2$ can be expressed in the standard way\cite{Ef:82b,SLA}
in terms of the replica NL$\sigma$M:
\begin{equation}
\label{S2SM}
S_2(\o)=-\lim_{n\to 0}\frac{1}{n^2}\frac{\partial^2}{\partial\o ^2}Z_n(\o),
\quad
Z_n(\o)=\int\D Q\exp\left(-{\cal F}[Q;\omega]\right)
\end{equation}
where ${\cal F}[Q;\omega]$ is the non-linear sigma model functional
\begin{eqnarray}
\label{nlsm}
{\cal F}[Q;\omega]=\frac{1}{L^d}\int \text{d}^dr
\hbox{Tr}\left[\frac{1}8D(\nabla Q)^2-\frac{i\omega \alpha}4\Lambda Q\right]\,.
\end{eqnarray}
This functional has been derived \cite{EfLKh}
for the model of free electrons in a random potential represented in terms
of the fermionic (anticommuting)  replica fields. As a result, 
the field $Q(\bbox{r})$ is an
Hermitian matrix of rank $2n$ whose elements are quarternions
for the orthogonal symmetry ensemble,
 complex numbers for the unitary symmetry ensemble which 
arises when  time-reversal symmetry in the model is broken,  
and real numbers for the symplectic  symmetry ensemble,
which  arises when   spin-rotation symmetry is broken.
 It satisfies the standard constraints
\begin{eqnarray}
Q^2=\openone _{2n}\,,\qquad\quad \Tr Q=0\,.
\label{Q2=1}
\end{eqnarray}
These constraints are resolved by representing $Q$ as follows:
\begin{equation}
\label{UQU}
Q=U^\dag \L U=T^\dag \L T\,,\qquad \L =\text{diag} (\openone _n, -\openone _n) 
\,,\qquad 
T =\exp\left(\begin{array}{cc}0_{n}&t\\-t^\dag &0_n
\end{array}\right)\,.
\end{equation}
Here $t$ is an arbitrary $n\!\times\! n$ matrix, and
$T$ is obtained by factorizing matrices $U\in \S(2n)$ with respect
to redundant matrices $R\in\S (n)\times \S (n)$ that commute with $\L $,
i.e.\ 
$U=RT$, where $\S (n)$ is
the appropriate symmetry group.
These conditions mean that $Q$ belongs to the {\it compact} Grassmannian
manifold \cite{DFN} (coset space), $\S (2n)/\S (n)\times \S (n)$.
For three different classes of symmetry, 
 $\S (n)$ becomes the {\it symplectic} group, $\Sp (n)$,  for the 
orthogonal class,
the unitary group $\U (n) $ for the 
unitary class, and the  {\it orthogonal} group, 
$\SO (n)$, for the symplectic class,

The ergodic regime corresponds to the level separations much smaller
than the Thouless energy, $D/L^2$ (which in the chosen units coincides, up
to a numerical factor, with the dimensionless conductance $g$). 
In this regime the gradient term in Eq.~(\ref{nlsm}) may be neglected, and
the NL$\sigma$M functional reduces to the zero-dimensional limit:\cite{Ef:82b}
\begin{equation}
\label{0d}
{\cal F}[Q;\omega]=
-\frac{i\omega \alpha}4\Tr\left[ \Lambda Q\right]\,,
\end{equation}
with $Q$ becoming a spatially homogeneous matrix. Here and in Eq.~(\ref{nlsm})
$\alpha=1$ for the orthogonal class, and $\alpha=2$ for the unitary  and
symplectic classes.
This factor arises because unitary and symplectic classes 
have been obtained from orthogonal \cite{EfLKh}
by the suppression of massive modes
corresponding to the time-reversal
or spin-rotational symmetry breaking and a subsequent  
reduction of the $Q$ matrix rank.
 The coefficient $\alpha$ also absorbs an extra factor 
in the symplectic case due to the redefinition
of the mean level spacing $\d $ in the chosen units.

\section{Unitary Ensemble}\setcounter{equation}{0}

\label{sec:unitary}

We limit all further considerations to the ergodic regime only, and
rewrite explicitly $Z(\o )$ given by Eqs.\ (\ref{S2SM}) and
(\ref{0d}):
\begin{equation}
\label{UZ}
Z_n(\o)=\int\D Q\exp\left[-i\frac{\o}{2}\Tr\L Q\right]\,,
\end{equation}
where the measure is defined by
\begin{equation}
\label{mes}
\D Q = \prod_{i,j=1}^n d\Omega _{ij}^{ra}\, d\Omega _{ij}^{ra\,*}\,,
\qquad d\Omega  \equiv dT\cdot T^{-1}\,.
\end{equation}
Here $T$ is the matrix parameterizing $Q$, Eq.~(\ref{UQU}), and 
$r$ and $a$ refer to the replica indices which originate from $G^r$ and
$G^a$, respectively.  
In the large-$\o $ limit this integral is mainly contributed
by the extrema of the functional which obey
the standard condition $ [\L, Q]=0$. This condition is
satisfied by any matrix of the form
$Q=\text{diag}(Q^r, Q^a)$,  where $Q^{r}$ and $Q^{a}$ 
 are the $n\times n$ Hermitian
matrices whose eigenvalues are $\pm1$ and $\Tr (Q^r+ Q^a)=0$.
This defines a highly degenerate saddle-point manifold which consists of
$C_{2n}^n$ submanifolds specified by a particular distribution 
of $n$ eigenvalues `+1' and $n$ eigenvalues `-1' between $Q^{r}$ and $Q^{a}$.
These submanifolds can be divided into
$n+1$ classes of equivalence, $Q_p=\text{diag}(Q^r_p, Q^a_p)$, labeled by
$\Tr Q^r_p=-\Tr Q^a_p=n-2p$, with $p=0,1,\ldots n$. The $p$-th class 
has weight $(C_{n}^p)^2$, with $C_{n}^p\equiv {n \choose p}$. 

The matrix $ Q^r_p$ with $(Q^r_p)^2=\openone_n$ and $\Tr Q^r_p=n\!-\!2p$
can be parameterized by analogy with Eq.~(\ref{UQU}) as
\begin{equation}
\label{Qrp}
 Q^r_p=( T^r_p)^\dag \l _p  T^r_p \,, \qquad
\l _p\equiv \text{diag}( {1 } _{n-p}, -1 _p)\,, \qquad
  T^r_p =\exp\left(\begin{array}{cc}0_{n-p}&t^r\\-(t^r)^\dag &0_p
\end{array}\right)\,,
\end{equation}
where $t^r$ is an arbitrary $p\times (n\!-\!p)$ matrix. 
This defines
the coset space $\textsf{G}_p=\U(n)/\U(n\!-\!p)\!\times\! \U(p)$, i.e.\ 
Therefore,
$Q_p = \text{diag}(Q^r_p, Q^a_p)$ belongs to the manifold $\textsf{G}_p\times
\textsf{G}_p$ and can be parameterized
as 
\begin{equation}
\label{Qp}
Q_p=T_p^\dag \L _p T_p\,,\qquad T_p=\text{diag}(T^r_p, T^a_p)
\,,\qquad
\L _p = \text{diag}(\l _p, -\l_p)
\end{equation}
The integer $p$ specifies the replica-symmetry breaking, as it describes
the number of the $-1$ eigenvalues in each $Q^r$ block (equal to the number of
the $+1$ eigenvalues in each $Q^a$ block): in the symmetry-unbroken case,
$p=0$, and hence retarded and advanced blocks, $Q^{r,a}$, contain only positive
or negative eigenvalues, respectively. 
We want to emphasize that manifolds $\textsf{G}_p\times
\textsf{G}_{p'}$ with
$p\ne p'$ cannot appear within the degenerate saddle-point manifold, 
$[\L, Q]=0$, of the functional (\ref{0d}). Indeed, 
the corresponding matrices have non-zero trace, and thus do not
belong to the $Q$ space defined by Eq.~(\ref{UQU}). Naturally, one could
have derived the replica NL$\sigma$M with different numbers of elements corresponding
to the retarded and advanced blocks, as in Ref.~\onlinecite{Weg:79}. In this
case, the replica-symmetry breaking would still mean the redistribution
of $+1$'s and $-1$'s eigenvalues between these blocks, keeping the
value of $\Tr Q$ the same (equal to the difference between the numbers
or replicas in the retarded and advanced blocks in the the
symmetry-unbroken case, $p=0$) in each $p$-th manifold.

Now we need to take into account contributions from
`massive' modes (with mass $\propto 1/\o $)
in the  vicinity of each manifold (\ref{Qp}).
In the large-$\o $ limit these contributions may be
considered as independent and the partition function is then represented by
the sum of all of them:
\begin{equation}
\label{Z}
Z_n(\o)=\sum_{p=0}^{n}\left(C_n^p\right)^2
\int\D Q \exp\left[-i\frac{\o}{2}\Tr \L_p Q \right],
\end{equation}
Here we have used the fact
that the similarity transformation in the vicinity
of $\L _p$, i.e.\ $U\L _p U^\dag$, covers the entire symmetric manifold
of the NL$\sigma$M, Eq.~(\ref{UQU}), including all the massive modes. 
Having substituted this into the functional (\ref{0d}),
we have reduced $\Tr U \L _p U^\dag \L$ to $\Tr \L _p T^\dag \L T=
\Tr \L _p Q$, where we have substituted
 $U=RT$, as defined after Eq.~(\ref{UQU}). 
Let us emphasize again that the above representation of $Z$ as the sum
over all $p$ 
can be justified only as a perturbative (in $1/\o $) procedure: a possible 
overlapping of massive modes originated from different manifolds is 
irrelevant in the large-$\o$ limit.

Each term in the sum ~(\ref{Z}) contains both massive and massless modes. 
Indeed, we have used above the factorization $U=RT$
with $R$ being block-diagonal matrices commuting with $\L $. The matrices
$T$ in $\Tr \L _p Q=\Tr \L T \L_p T^\dag$ still contain the subset
of matrices commuting with $\L _p$ that correspond to the massless modes. 
Therefore, we need to parameterize $T$ in a way which enables us to factorize
out these massless modes and perform the integration over the massive ones. 

The most suitable parameterization of $T$,
 analogous to that used in Ref.~\onlinecite{VZ:85}
for the bosonic NL$\sigma$M, can be obtained 
by expanding the matrix exponent in Eq.~(\ref{UQU}). By introducing 
matrix $B\equiv t ({t^\dag t})^{-1/2} {\sin\sqrt{t^\dag  t}}$,
we represent $T$ and thus $Q=T^\dag \L T$ as follows:
\begin{equation}
\label{p1}
T =
\left(
\begin{array}{cc}
\displaystyle
 \sqrt{\openone_n-BB^\dag }  &
 \displaystyle
 B  \\
\displaystyle
-B^\dag  &
\displaystyle
\sqrt{\openone_n-B^\dag B}
\end{array}
\right)\,,
\qquad
Q = \left(
\begin{array}{cc}
\openone_n -2BB^\dag  & B\sqrt{\openone_n -B^\dag B} \\
B^\dag \sqrt{\openone_n -BB^\dag } & -(\openone_n -2B^\dag B)
\end{array}\right).
\end{equation}
The matrix $B$ in this parameterization is not unconstrained, though. The
$Q=Q^\dag$ condition is fulfilled only when the matrices
$\sqrt{\openone_n -BB^\dag }$ and $\sqrt{\openone_n -B^\dag B}$ are Hermitian.
This is so only when
all the  eigenvalues of $BB^\dag $ and $B^\dag B$ do not exceed
unity. Only under this constraint does 
$Q$, parameterized as in Eq.\ref{p1}, still belong to the
coset space $\U(2n)/\U(n)\times \U(n)$.
Nevertheless, this parameterization is very convenient. 
 First, the corresponding Jacobian is unity
(see Appendix), so that the measure of integration (\ref{mes}) can be
written simply as
\begin{equation}
\label{BB}
\D Q = \prod_{i,j}dB_{ij}dB^*_{ij}\equiv \D B\,.
\end{equation}
In addition, the representation of 
all the exponents in the sum~(\ref{Z}) in terms of $B$ is also very simple,
$
\Tr \L _p Q=2(n-2p) - 2\Tr 
 \l_p (BB^\dag  + B^\dag B)
$, so that we obtain: 
\begin{eqnarray}
\label{int}
Z_n(\o)&=&\sum_{p=0}^{n}\left(C_n^p\right)^2\cdot e^{i\o(2p-n)}\,
Z_n^p(\o)\,,\\
Z_n^p(\o)&=&
\int \!\D B
\exp\left[i\o\Tr \l_p (BB^\dag  + B^\dag B) \right].
\label{Znp}
\end{eqnarray}
The region of integration in (\ref{Znp})
is restricted by the constraint described 
after Eq.~(\ref{p1}).
Last, but not least, the parameterization (\ref{p1}) allows us to separate out
the massless modes, which obey the condition $[T, \L _p]=0$,
in each integral (\ref{Znp}). Indeed, this 
condition is satisfied by all matrices $T$ constructed from $B$ which 
anticommute with $\l _p$, i.e.\ have the off-diagonal block structure. 

This means that in the 
representation of $B$ in the block form
reflecting the structure of $\l_p={\rm diag}(\openone_{n-p}, -\openone_p)$,
\begin{equation}
B = \left(
\begin{array}{cc}
  B_1   &  b_1 \\
 b_2^\dag    &  B_2
\end{array}\right),
\label{Bb}
\end{equation}
the matrices $B_{1,2}$ represent the massive modes, and $b_{1,2}$ massless. 
When the massive modes are suppressed ($B_1=0$ and 
$B_2=0$), the $T$ matrices in Eq.~(\ref{p1}) constructed from
$p\times(n\!-\!p)$ matrices $b_{1,2}$ 
only, parameterize the same degenerate $p$-the manifold, $\textsf{G} _p
\!\times\! \textsf{G} _p$ described in Eq.~(\ref{Qp}), as one expects.

By substituting the representation (\ref{Bb}) into
Eq.~(\ref{Znp}), we reduce $Z_n^p$ to the product of integrals over
the massive and massless modes:
\begin{equation}
\label{comp}
Z^p_n(\o)=
\int \D B_1\, \D B_2 \exp\left[-2i\o\tr(B_1 B_1^\dag  - B_2 B_2^\dag )\right]
\int \D b_1 \D b_2 ,
\end{equation}
Here the region of
integration over $b_{1,2}$ depends on $B_{1,2}$ due to the constraint
on the eigenvalues of the matrices $B B^\dag $ and
$ B^\dag B$ in the representation (\ref{Bb}).
 Since the 
integral over $B_{1,2}$ is contributed only by
the region where both $\tr B_1 B_1^\dag $ and
$\tr B_2 B_2^\dag  
\alt1/\o \ll1$,
in the leading in $1/\o$ approximation we may
put both $B_{1,2}$ to $0$ in the constraint of the integration 
 region over the massless modes $b_{1,2}$. In this approximation, as we
have noticed after Eq.~(\ref{Bb}), matrices $b_{1,2}$ parameterize the
$p$-th manifold (\ref{Qp}) so that
\begin{equation}
\label{bb}
\int \D b_1 \D b_2 =\int \D Q_p=\Omega^2(\textsf{G}_p)\,,
\end{equation}
where the measure of integration over $\D Q_p$ is defined in terms of $T_p$
in the same way that $\D Q$ is defined in terms of $T$, Eq.~(\ref{mes}), 
and $\Omega(\textsf{G}_p)$ is the volume 
of the compact coset space $\textsf{G}_p$. This volume is expressed 
via the well-known volumes of the unitary group, $\Omega(\U (n))$, as follows:

\begin{equation}
\label{vu}
\Omega(\textsf{G}_p)=\frac{\Omega(\U(n))}{\Omega(\U(n-p))\Omega(\U(p))}=
(2\pi)^{\frac{1}{2}[n^2-(n-p)^2-p^2]}\prod\limits_{j=1}^{p}
\frac{\Gamma(1+j)}{\Gamma(n + 2 - j)}.
\end{equation}
In the same large-$\o $ approximation, 
 the variables $B_{1,2}$ parameterizing the massive modes
are unconstrained. Then the Gaussian integral over the $2[(n\!-\!p)^2 + p^2]$
independent massive modes yields
\begin{equation}
\label{GI}
\tilde Z^p_n(\o)\equiv
\int \D B_1\, \D B_2 \exp\left[2i\o\tr(B_1 B_1^\dag  - B_2 B_2^\dag )\right]=
\left(\frac\pi{-i\o } \right)^{(n-p)^2} 
\left(\frac\pi{i\o } \right)^{p^2} 
\end{equation}
Combining Eqs.~(\ref{GI}) and (\ref{vu}) and omitting
 an irrelevant overall factor which goes to $1$
when $n\to0$, we arrive at the following expression 
which is essentially the same as that 
derived in Ref.~\onlinecite{KamMez:99a} via the Itzykson-Zuber integral:
\begin{equation}
\label{un}
Z_n(\o)=\sum_{p=0}^{\infty}\left[F_n^p\right]^2\cdot
\frac{e^{i\o(2p-n)}}{(2\o)^{(n-p)^2+p^2}}, \qquad
F_n^p\equiv C_n^p \prod\limits_{j=1}^{p}\frac{\Gamma(1+j)}
{\Gamma(n+2-j)}.
\end{equation}
Here the summation over $p$ has been extended to $\infty$ since 
$F^p_n=0$ for all integer $n>p$. This allows one to take
the replica limit, $n\to0$, in each of the terms in Eq.~(\ref{un}).
Due to the fact that $F^p_n\propto n^p$ as $n\to 0$, 
only the terms with $p=0$ and $p=1$ 
 in Eq.~(\ref{un}) contribute to $S_2$ in Eq.~(\ref{S2SM}).
Omitting all the terms with $p\ge2$, one obtains
\begin{equation}
\displaystyle
Z_n(\o)=\frac{e^{-i\o n}}{\o^{n^2}} +
n^2 \frac{e^{i\o(2-n)}}{4\o^{(n-1)^2+1}}\,.
\end{equation}
Substituting this into Eqs.~(\ref{S2SM}) and (\ref{S2}) and keeping
the leading in $1\o $ terms only, one arrives
at the expression (\ref{R2U}) for the TLCF.
Although this expression looks as being exact, it has been actually derived
only in the large-$\o $ limit, as is the case  in the Itzykson-Zuber
calculation of Kamenev and Mezard \cite{KamMez:99a}, and in the 
`supersymmetry breaking' method  of Andreev and Altshuler. \cite{A2}

\section{Orthogonal and Symplectic Ensembles}\setcounter{equation}{0}

\label{sec:ort}

A generalization to both these symmetries from the unitary one is
straightforward:
elements of all matrices in the previous section were complex numbers, while 
now they become quarternion numbers \cite{EfLKh,DFN} in the orthogonal 
symmetry and real numbers in the symplectic symmetry. In other words,
$\S(n)$, $\S(p)$, etc,
should be substituted for $\U (n)$, $\U (p)$, etc,  in all formulae of
Section~\ref{sec:unitary}, with 
$\S$ being the symplectic group $\Sp $ for the orthogonal symmetry
and the orthogonal group $\SO$ for the symplectic symmetry (such an
inversion is characteristic of the fermionic replica approach). 
One should also redefine the measure of integration, Eq.~(\ref{mes}), 
in terms of independent matrix elements and substitute $2\o $
by $\alpha \o$ as defined in Eq.~(\ref{0d}).

After these redefinitions,  the partition function $Z_n$ 
is represented by the sum (\ref{int}), with each term in this sum
given by the  product of integrals over
the massive and massless modes, Eq.~(\ref{comp}).
The number of massive modes is  $(4/\beta)[(n-p)^2+p^2]$,
where $\beta =1,\,2\,4$ for the orthogonal, unitary and  symplectic symmetry
is the standard Dyson parameter. For  orthogonal and symplectic symmetries,
the Gaussian integral (\ref{GI}) over the
massive modes yields 
\begin{equation}
\label{GI2}
\tilde Z^p_n(\o)=\cases{
\displaystyle \left(\pi/\o \right)^{2[(n\!-\!p)^2 + p^2]}\quad
(-1)^{(n\!-\!p)^2 - p^2}& orthogonal
symmetry\cr
\displaystyle \left(\pi/2\o \right)^{[(n\!-\!p)^2 + p^2]/2}\;
i^{-[(n\!-\!p)^2 - p^2]/2}& symplectic
symmetry}
\end{equation}
 The integrals over massless modes, Eq.~(\ref{bb}), give
the volumes, $\Omega^2(\textsf{G}_p^\beta)$ of the corresponding coset spaces.
The results can be written as
\begin{equation}
\label{ml}
 \Omega(\textsf{G}_p^\beta) =\cases{
\displaystyle (2\pi )^{2(n\!-\!p)p}\;
\prod\limits_{j=1}^{p}
\frac{\Gamma(1+{2}j)}{\Gamma[1+{2}(n+1-j)]}& orthogonal
symmetry\cr
\displaystyle (\pi )^{(n\!-\!p)p/2}\;
\prod\limits_{j=1}^{p}
\frac{\Gamma(1+j/{2})}{\Gamma[1+(n+1-j)/2]}& orthogonal
symmetry}
\end{equation}
Combining Eqs.~\ref{GI2} and \ref{ml},  we obtain the partition 
function for both the symmetries. It can be written in terms of
$\beta$, together
with the unitary $Z_n$, Eq.~(\ref{un}), in universal form:
 
\begin{equation}
\displaystyle
Z_n(\o)=\sum_{p=0}^{\infty}F^2_{\beta}(n,p)
\frac{e^{-i\o (2p-n)}}{(2\o)^{\frac{2}{\beta}[(n-p)^2+p^2]}}, \quad
F_{\beta}(n,p) = C_n^p \prod\limits_{j=1}^{p}
\frac{\Gamma(1+\frac{2}{\beta}j)}{\Gamma(1+\frac{2}{\beta}(n+1-j))}.
\end{equation}
In the $n\to 0$ limit
only the terms with $p=0,1$ survive for $\beta=1,\,2$, 
while for $\beta=4$ one needs to keep the $p=2$ term as well:
 
\begin{equation}
\displaystyle
Z_{n\to 0}(\o)=1-i\o n+n^2\left\{
-\frac{\o^2}{2}-\frac{2}{\beta}\ln\o +
\frac{\Gamma^2(1+2/\beta)}{(2\o)^{\frac{4}{\beta}}}\cdot e^{2i\o}
+ \delta_{\beta,4}\frac{e^{4i\o}}{2^8\o^4}.
\right\}
\end{equation}
Substituting this into Eqs.~(\ref{S2SM}) and (\ref{S2}), we obtain the TLCF
as follows:
\begin{equation}
\displaystyle
R_2(\o) = -\frac{1}{\beta\o^2}
+ \frac{2\Gamma^2(1+2/\beta)}{(2\o)^{\frac{4}{\beta}}}\cos 2\o
+ \delta_{\beta,4}\frac{\cos 4\o}{32\o^4}.
\end{equation}
This reproduces the correct large-$\o$ asymptotic behavior of
$R_2$ for all three ensembles, which includes the non-perturbative
oscillatory factors. 

As we have derived these results with a non-standard treatment of 
the standard fermionic NL$\sigma$M, they also should be contained in 
the exact integral representation obtained for this model 
by Verbaarschot and Zirnbauer. \cite{VZ:85} 
We will show now that this is, indeed, the case. 
 
\section{The large-{\large $\o $} limit of the Verbaarschot-Zirnbauer Integral}
\label{sec:VZ}\setcounter{equation}{0}

The `zero-mode' partition  function, Eq.~(\ref{0d}), can be exactly
represented in the following form
\begin{equation}
\label{VZ}
Z_n(\o)=\Omega^2(\U(n))\int_{-1}^{+1} \d^2(\l)\,
\prod_{i=1}^{n}
e^{-i\o\l_i}\text{d}\l_i \,, \qquad \d(\l) \equiv \prod_{i<j}(\l_j - \l_i)\,.
\end{equation}
which is equivalent (with accuracy up to factors going to $1$ in the 
$n\to0$ limit) to the representation for $S_2$ given in 
Eq.~(2.24) of the 
paper by  Verbaarschot and Zirnbauer. \cite{VZ:85}

The leading in $1/\o$  contributions
to this highly oscillatory integral (which does not have stationary
points inside integration region) come from the end points. To single
out  these contributions, we must take some $\l$'s close to $+1$ and the rest
close to $-1$, which imitates replica symmetry breaking.
Let us choose $n\!-\!p$ of $\l $'s  close to $+1$ and $p$ of $\l $'s 
close to $-1$. Then we can  split up the
Vandermonde determinant in the following way:
$$
\d^2(\l ) =\prod_{i,j}^n|\l_j - \l_i|\approx
2^{2p(n\!-\!p) } \Delta^2_+\,\Delta^2_- 
$$
where
 
\begin{equation}
\Delta^2_+ = \prod_{i,j=1}^{n-p}|\l_i-\l_j|\, \quad
\Delta^2_- = \prod_{i,j=n-p+1}^{n}|\l_i-\l_j|.
\end{equation}
Reducing the integral (\ref{VZ}) to the sum of such contributions only, 
we represent it as
 
\begin{equation}
Z_n(\o)\approx\Omega^2(\U(n))\sum_{p=1}^{n}(C_n^p)\,2^{2p(n-p)}
\int_{-\infty}^{+1} \Delta^2_+\, \prod_
{j=1}^{n-p} \ee ^{-i\o\l_j}\,\dd \l_j\;
\int_{-1}^{+\infty}\Delta^2_- \!\!  \! \prod_
{j=n-p+1}^n \ee ^{-i\o\l_j}\,\dd \l_j\,.
\end{equation}
Since in each of the integrals all the variables are close to one of 
the limits of integration, the second limit was extended to
infinity. 
Now we make substitutions
 $\l_i = 1-x_i$ in the first integral,  and $\l_i = -1+x_i$ in the
second one, reducing the above sum to the form:
\begin{equation}
\label{sum}
Z_n(\o)\approx\Omega^2(\U(n))\sum_{p=1}^{n}(C_n^p)\,2^{2p(n-p)}
\ee ^{i\o (2p\!-\!n)} I_{n\!-\!p} I_p
\end{equation}
where $I_p$ are integrals of the Selberg's type: \cite{RMT}

\begin{equation}
I_p=\int\limits_{0}^{\infty}\Delta^2(x)\prod_{j=1}^{p}\text{d}x_j 
\text{e}^{-i\omega x_j}
\end{equation}
Substituting the known Selberg integrals and
 discarding an overall factor
which goes to unity in the replica limit we arrive at
\begin{equation}
\displaystyle
Z_n(\o)=\sum_{p=0}^{n}\left[F_n^p\right]^2\cdot
\frac{e^{i\o(2p-n)}}{2\o^{(n-p)^2+p^2}}, \quad
F_n^p= C_n^p \prod\limits_{j=1}^{p}\frac{\Gamma(1+j)}
{\Gamma(n+2-j)}.
\end{equation}
This expression is exactly the same as Eq.~\ref{un} obtained in 
Section~\ref{sec:unitary} with the help of the replica-symmetry breaking. 
Therefore, the exact representation (\ref{VZ}) {\it does contain}
the true oscillatory asymptotic behavior of the TLCF.

The authors of Ref.~\onlinecite{VZ:85}
have also drawn attention to the fact that 
there is an apparent contradiction between the $\o =0$ limit for 
$S_2$ obtained from the replica trick and the exact supersymmetric result.
Indeed, if $\o$ is put to $0$ in the expression for $S_2$ following from
Eqs.~(\ref{VZ}) and (\ref{S2SM}) the $n\to0$ limit is taken 
{\it after that}, one obtains $S_2(\o\!=\!0)=-1$.  
This cannot be correct 
as $\Re\text{e}\, S_2(\o\!\to\! 0)>0$ as follows from the definition
(\ref{S2}) and, moreover, it is known that 
$ S_2(\o\!\to\! 0)\to\delta(\o)$. 
What is interesting, however,  is that if one separates
 the  singular, $S_2^{\text{sing}}(\o)$, and
regular, $S_2^{\text{reg}}(\o)$, parts of $S_2(\o)$,  then 
$S_2^{\text{reg}}(\o\!\to\! 0)=-1$.
Therefore, the replica method gives $S_2^{\text{reg}}(\o\!=\!0)$
correctly, and it is just 
$S_2^{\text{sing}}$  which is missing.
However, the fact that $Z_n(\o \!\to\!0)$ is finite for any integer $n$ does not
necessarily implies that it is also finite (as a function of $\o\to0$)
in the replica limit. 
For example, if the expansion of $S_2 (\o )$ {\it before} taking the $n=0$
limit
contained a term proportional to $\o^{n^2-1}$, it would be 
singular in the replica limit. In other words, if a non-trivial
dependence on the order of limits $n\to0$ and $\o\to0$ existed,
in a spirit of the replica trick the $n\to0$ limit should be taken first.
At the moment, though, this is only a speculation as we have not yet succeeded 
 in calculating the integral (\ref{VZ}) for small $\o $.
However, the fact that the large-$\o $ limit of this integral reproduces the
correct results (\ref{R2U}) makes it plausible that there are only technical
difficulties rather than one of principle
 in the application of the replica method. 
 
\section{Conclusions}
\label{sec:con}

The purpose of this paper was to demonstrate explicitely that non-perturbative
oscillatory contributions to the TLCF of electrons in a random
potential could be extracted from
 the standard NL$\sigma$M
formulated in fermionic replicas by Efetov, Larkin and Khmelnitskii\cite{EfLKh}
many years ago.
To this end, all one needs is to parameterize all the non-trivial saddle-point
manifolds corresponding to the broken replica-symmetry 
and describing `massless modes' of the theory, and expand the action 
in the vicinity of these manifolds to include `massive modes'.
The very sumilar approach has been used in the supersymmetric NL$\sigma$M:
 the non-perturbative oscillations
have been extracted by the expansion around two extremal points one of which
breaks the supersymmetry. \cite{A2} 
Since the exact supersymmetric calculation of the TLCF was well known, 
\cite{Ef:82b} it
was clear that the supersymmetry breaking\cite{A2}
was just a convenient method 
of extracting the large-$\o $ limit (and going beyond  the universal
`zero-mode' apporximation). It is not clear at the moment whether exact
calculation is possible within the replica approach. 
However, as we have shown in the Section \ref{sec:VZ}, 
the exact integral representation of $R_2(\o )$ does at least describe the 
correct behavior in the large-$\o $ limit.

On the other hand, there is a number of non-perturbative
results where all the leading  contributions come from the 
small $\o $ limit (see, e.g., Ref.~\onlinecite{KhM}):
the existence of a nontrivial spatially inhomogeneous saddle point
in the non-compact sector of  the supersymmetric NL$\sigma$M
was the main source of these contributions. 
At the moment it is unclear whether
such a  saddle point could be found within the fermionic
replica approach.
Nevertheless, the fact that some of these results had earlier been obtained
\cite{AKL:86,AKL:91} within the RG treatment
of the fermionic replica  NL$\sigma$M  prompts that it could be  
technical rather than methodoligical difficulties involved.

Despite the fact that the replicas prove now to be a trusted tool for 
non-perturbative calculations, there is no doubt that the
supersymmetric NL$\sigma$M is the best tool to obtain non-perturbative
results 
for non-interacting electrons.
But the further elaboration of the non-perturbative
methods within the replica trick will be important for the problems where
supersymmetry cannot be applied.  This is the situation with the interacting 
electrons in a random potential. Such systems can be described by NL$\sigma$M
with fermionic replicas only and we believe that the further development of
methods similar to  presented here  will enable one to obtain at last 
non-perturbative results for disordered systems of interacting electrons.

\acknowledgments
This work has been supported by the
EPSRC grant GR/K95505.

\appendix
\def\theequation{A.\arabic{equation}}%

\section{Calculation of Jacobian }

To calculate differentials of matrices  entering the  measure (\ref{mes})
we use the following rational substitution:
\begin{equation}
\label{rp}
\displaystyle
B= \frac{2}{1-\tau\tau^\dag }\,\tau 
\quad \Longrightarrow \quad
T=\left(
\begin{array}{cc}
\displaystyle
 \frac{1+\tau\tau^\dag }{1-\tau\tau^\dag } &  
 \displaystyle
 \frac{2}{1-\tau\tau^\dag }\,\tau   \\
\displaystyle
-\frac{2}{1-\tau^\dag \tau}\,\tau^\dag  & 
\displaystyle
\frac{1+\tau^\dag \tau}{1-\tau^\dag \tau}
\end{array}
\right).
\end{equation}

Then the differentials in the measure  (\ref{mes})
are given by

\begin{equation}
\label{d1}
\dd \Omega^{12}(T)\equiv \left(\dd T\cdot T^\dag \right) =
2\frac{1}{1-\tau\tau^\dag }\left(\dd \tau -\tau\,\dd \tau\,\tau\right)
\frac{1}{1-\tau^\dag \tau}.
\end{equation}
On the other hand we can find the differential of $B$, entering the measure
(\ref{BB}) as follows:
\begin{equation}
\label{d2}
\dd B=2\frac{1}{1-\tau\tau^{\dag} }\left(\dd \tau +\tau\,\dd \tau\,\tau\right)
\frac{1}{1-\tau^\dag \tau}.
\end{equation}
Comparing Eqs.~(\ref{d1})
and (\ref{d2}) one can see that the Jacobians of transitions
$\dd \Omega^{12}(T)
\longrightarrow \dd \tau$ and $\dd B\longrightarrow \dd \tau$ are exactly
the same. From that fact one deduces that the Jacobian of the transition
$\dd \Omega^{12}(T) \longrightarrow \dd B$ is unity.


\end{document}